\documentclass[aps,pre,twocolumn,superscriptaddress,showpacs] {revtex4}

\usepackage{amsfonts}
\usepackage{amsmath}
\usepackage{fontenc}
\usepackage[latin1]{inputenc}
\usepackage{dcolumn}
\usepackage{bm}
\usepackage{graphicx}

\begin{document}

\title{Number of propagating modes of a diffusive periodic waveguide in the semiclassical limit}

\author{Felipe Barra}
\affiliation{Departamento de F\'isica y CIMAT, Facultad de Ciencias F\'isicas y Matematicas, Universidad de Chile, Casilla 487-3, Santiago Chile}
\author{Agnes Maurel}
\affiliation{Laboratoire Ondes et Acoustique, UMR CNRS 7587, Ecole Sup\'erieure de Physique et Chimie Industrielles, \\10 rue Vauquelin, 75005 Paris, France}
\author{Vincent Pagneux}
\affiliation{Laboratoire d'Acoustique de l'Universit\'e de Maine, UMR CNRS 6613 \\Avenue Olivier Messiaen, 72085 Le Mans Cedex 9 ,France}
\author{Jaime Zu\~niga}
\affiliation{Departamento de F\'isica, Facultad de Ciencias F\'isicas y Matematicas, Universidad de Chile, Casilla 487-3, Santiago Chile}


\date{\today}

\begin{abstract}
We study the number of propagating Bloch modes $N_B$ of an infinite periodic billiard chain. The asymptotic semiclassical behavior of this quantity depends on the phase-space dynamics of the unit cell, growing linearly with the wavenumber $k$ in systems with a non-null measure of ballistic trajectories and going like $\sim \sqrt{k}$ in diffusive systems. We have calculated numerically $N_B$ for a waveguide with cosine-shaped walls exhibiting strongly diffusive dynamics. The semiclassical prediction for diffusive systems is verified to good accuracy and a connection between this result and the universality of the parametric variation of energy levels is presented.
\end{abstract}

\pacs{05.45.Mt, 03.65.Sq, 42.25.Bs, 05.45.Pq, 05.60. -k}

\maketitle 

\section{Introduction}
According to the Bohigas-Giannoni-Schmit conjecture \cite{Bohigas:1984p306}, a system with chaotic phase-space dynamics is expected to show universal quantum level statistics in the semiclassical limit,
consistent with the predictions of Random Matrix Theory (RMT). For instance, the two-point level correlation of closed chaotic systems has been shown experimentally and numerically to agree with the Gaussian Orthogonal Ensemble (Gaussian Unitary Ensemble) for systems with (without) time-reversal symmetry \cite{Stockmann:1999p473,Haake:2001p476}. Analytical results supporting this universality have also been achieved using the semiclassical trace formula, first by means of the diagonal approximation and more recently including non-diagonal terms verifying full agreement of the semiclassical form factor with 
RMT \cite{Heusler:2007p222}. \\

Much less is known for spatially extended chaotic systems. In the case of systems that relax by a diffusion process, the spectral properties deviate from random matrix theory in a particular way.  
Dittrich et al. \cite{Dittrich:1997p106, Dittrich:1998p118} considered the
statistical properties of the energy bands $E_{n,\theta}$ 
of a closed periodic diffusive system with $N$ chaotic unit cells and analyzed the signatures of chaotic diffusion in the form factor (the Fourier transform of the two-point level correlation function). They showed that their semiclassical results agree nicely with those of Simons and Altshuler \cite{Simons:1993p701,Simons:1993p111} who proposed that the fluctuations of the energy bands $E_{n,\theta}$ or in fact any parametric variation of the energy levels are universal within the appropriate Dyson ensemble (depending on time-reversal symmetry). Simons and Altshuler calculated the universal two-point level correlation function for a disordered system  under the influence of an external perturbation --for instance in the form of an Aharonov-Bohm flux--  and showed that the response of the spectrum to those parameter variations was universal after a suitable scaling with two system specific parameters. Dittrich et al. \cite{Dittrich:1997p106, Dittrich:1998p118} identified one of these parameters with the diffusion coefficient of the chain.\\

These correlation functions or form factors provide a detailed description of the statistical properties of the spectrum. In this work we consider a particular quantity directly related to the correlation function and of great practical importance for wave propagation in periodic media: the number of propagating modes $N_B(E)$ with energy $E$.
In a periodic system the propagating modes are given by the Floquet-Bloch modes, 
which are quasi-periodic solutions of the Schrodinger equation. They propagate ballistically through the system with constant (group) velocity and are the only modes allowed to transport energy. For instance, the dimensionless Landauer's conductance of a long periodic sample fluctuates closely to the number of these modes propagating in a given direction. In \cite{Faure:2002p3}, Faure studied this quantity for quasi-one-dimensional periodic systems in the small Plank constant limit $h=2\pi\hbar\rightarrow 0$. He showed that if the phase-space displays tori in the transverse stroboscopic Poincare section, corresponding to trajectories that propagate ballistically through the system, then $N_B\sim h^{-1}$ but, on the other hand, for fully chaotic and diffusive systems $N_B\sim h^{-1/2}$. \\

We study $N_B(E)$ numerically for periodic waveguides with diffusive 
classical dynamics and finite horizon in the semiclassical regime.
The paper is organized as follows. In Section \ref{section-bloch_modes} we review Faure's derivation of 
the number of propagating Bloch modes in a waveguide. In section \ref{Sec.3} we define the billiard where the predictions are tested. The results are summarized in section \ref{section-numerics}. Then in section \ref{Sec.5} we discuss in further detail our results and its relation to previous works and present some conclusions.

\section{Number of Bloch modes for $\hbar\rightarrow 0$}\label{section-bloch_modes}

In order to be self-contained, we start reviewing the essential results of \cite{Faure:2002p3}. 
First, in an infinite periodic waveguide, the distribution of quantum velocities $P_\hbar(v,E)$ determines the number of propagating modes $N_B(E)$ with positive velocity [see Eq.(\ref{exact})]. Second, the asymptotic form of the average $\langle N_B(E)\rangle$ for $\hbar \to 0$, obtained using a semiclassical approximation of $\langle P_\hbar(v,E)\rangle$ [see Eq. \ref{nb_k_diff}].\\

Consider a periodic waveguide in the cartesian $(x,y)$ plane, with $x\in ]-\infty,+\infty[$ (the propagation axis) and $y$ bounded for every $x$ between the wave boundaries that we define as hard walls. The waveguide is composed of cells with length $L$. In the waveguide particles are free, hence Schrodinger's equation reduces to Helmholtz equation, i.e. 
\begin{equation}
 \label{helmholtz}
 (\nabla^2 + k^2)\psi(x,y) = 0
\end{equation}
with $\psi(x,y)$ satisfying Dirichlet boundary conditions at the walls. The reader can have in mind several physical contexts where this is valid such as optical, microwave or acoustic cavities or degenerate electron gases in mesoscopic devices. The classical analog of the quantum system under consideration is given by the free particle Hamiltonian $H(x,p_x,y,p_y)= (p_x^2+p_y^2)/2$ (we consider particles of unit mass), plus perfect elastic walls as the waveguide boundaries. The results we present below hold for general periodic Hamiltonian systems (not only billiards) where the confining potential in the transversal direction $y$ could be soft. Eq. (\ref{nb_k_diff}) is the general result for that case whereas the particular result for hard-wall billiards is given by Eq. (\ref{nb_kk_diff}) which is the expression we use in the rest of the paper.

\begin{figure}
 \centering
 \includegraphics[width=0.70\columnwidth]{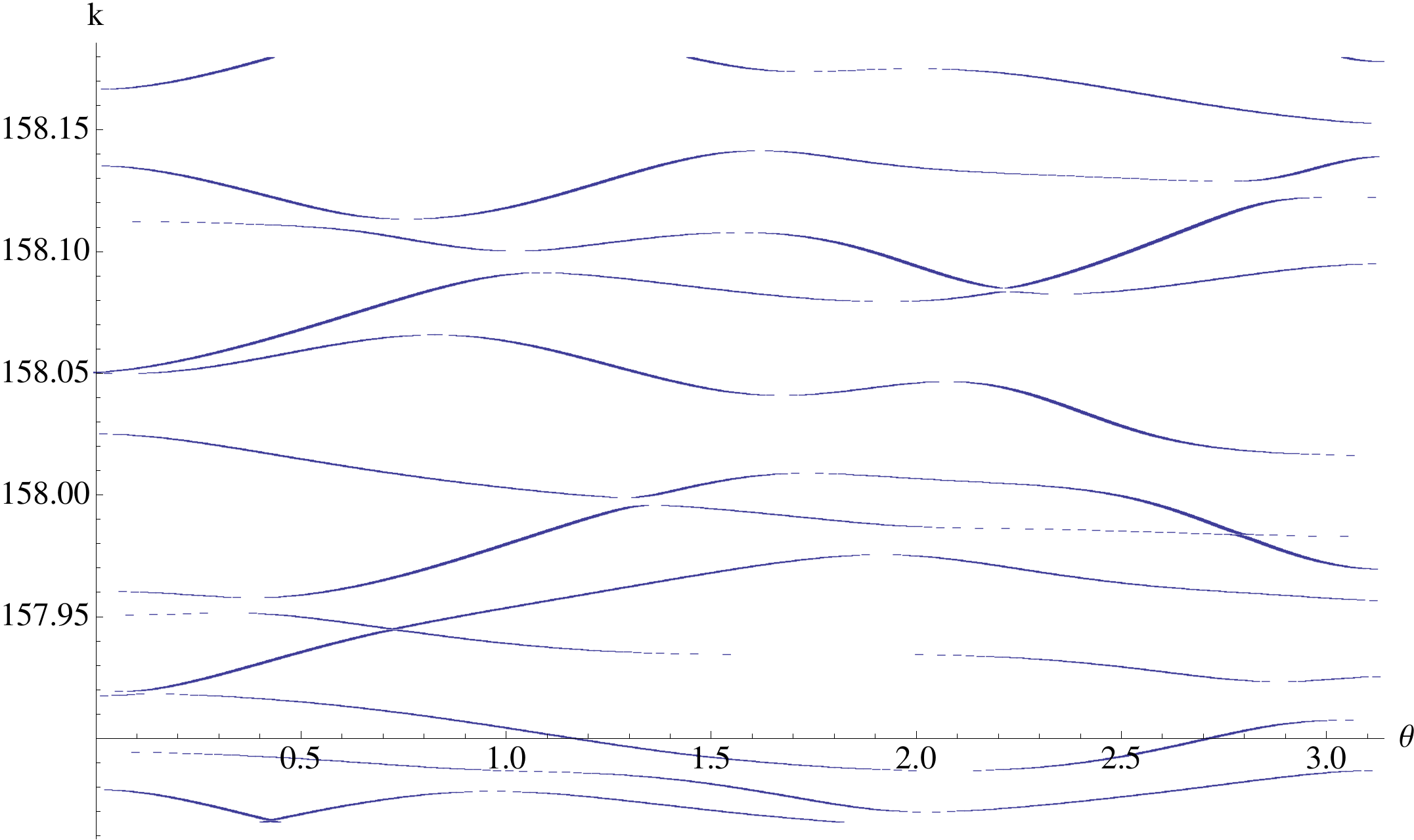}
 \caption{Part of the spectrum $k_n(\theta)$ for the billiard studied in Sec.\ref{section-numerics}. Note that for the statistics of $N_B(k)$ the computation consider values of $k$ considerably larger than the ones of this figure.}
 \label{bands} 
\end{figure}
    
Let us consider the phase-space surface of constant energy in one unit cell $\Sigma_E=\{ (x,p_x,y,p_y) \;/\; 0\leq x\leq L,\; h_1(x) \leq y \leq h_2(x), \; H(x,p_x,y,p_y)=E \}$ where $h_1(x)$ and $h_2(x)$ define the lower and upper walls of the waveguide respectively. Since the boundaries of the guide are periodic and $\Sigma_E$ is compact, we have from Bloch theorem that the solutions of Schrodinger's equation [Eq. (\ref{helmholtz})] form bands, i.e., its spectrum has the form $E_{n,\theta}$ with $n\in\mathbb{N}$ the band index and 
$\theta=q L\in [-\pi,\pi]$ where $q$ is the Floquet-Bloch wavenumber. Figure \ref{bands} illustrates this spectrum for a chaotic periodic waveguide.
An important consequence of Bloch theorem is that the propagation in a periodic system is ballistic with a group velocity \cite{kittel}
\begin{equation}
 \label{vh_def2}
v_{n,\theta}=\frac{L}{\hbar}\,\frac{d E_{n,\theta}}{d\theta}.
\end{equation}
Since $N= \int_{a}^{b} dx\, \delta(C-f(x)) \Theta(f'(x)) \frac{df}{dx}$ is the number of points such that 
$f(x)=C$ with positive derivative in the interval $(a,b)$, Faure expressed the number of Bloch modes with energy $E$ propagating with positive velocity, $N_B(E)$ as
\begin{equation}
N_B(E)= \sum_n \int_{-\pi}^{\pi} d\theta\, \delta(E-E_{n,\theta}) \Theta(v_{n,\theta}) \frac{dE_{n,\theta}}{d\theta}
\label{Nb_def2}
\end{equation}
where  $\delta(\cdot)$ is Dirac's delta function and $\Theta(\cdot)$ is Heaviside step function. Considering the distribution of quantum velocities $P_\hbar(v,E)$ defined as \cite{Asch:1998p100,Faure:2002p3}
\begin{equation}
 \label{Ph_def}
P_\hbar(v,E) =h^2 \sum_n \int \frac{d\theta}{2\pi}\, \delta(v-v_{n,\theta})\delta(E- E_{n,\theta}) 
\end{equation}
with $v_{n,\theta}$ defined in Eq.(\ref{vh_def2}) one obtains
\begin{equation}
 \label{exact}
 N_B(E)  = \frac{1}{h L} \int_0^\infty v P_\hbar(v,E) \,dv. 
\end{equation}
The normalization over $v$ of $P_\hbar(v,E)$ is 
\begin{equation}
\int dv P_\hbar(v,E)=h^2 \int \frac{d\theta}{2\pi}\rho(E,\theta)
\label{normalizacion}
\end{equation}
with $\rho(E,\theta)= \sum_n \delta(E- E_{n,\theta})$ the density of states. We note that Eq.(\ref{exact}) is an exact result, a particular case of the Kac-Rice formula \cite{wilkinson}, valid for quasi-one-dimensional periodic systems \cite{Faure:2002p3}.\\

The number of propagating Bloch modes is a rapidly fluctuating quantity as a function of $E$ at scales near the mean level spacing $\Delta E=1/\bar{\rho}_E$, where $\bar{\rho}_E$ is the mean density of states. These fluctuations are eliminated by considering the average of $N_B(E)$ over an energy interval $\delta E$ such that $E \gg \delta E \gg \Delta E$ (i.e., quantum mechanically large but classically small), which is the same definition used to determine $\bar{\rho}_E$ from $\rho(E,\theta)$. This procedure is supposed to play a similar role to the ensemble average for disordered systems. Note that we can write
\begin{equation}
\label{promedio}
 \langle N_B(E)\rangle  = \frac{1}{h L} \int_0^\infty v \langle P_\hbar(v,E) \rangle \,dv. 
\end{equation}
As follows from Eq.(\ref{normalizacion}), $\langle P_\hbar(v,E) \rangle$ is normalized to $h^2\bar{\rho}_E=\nu_E$ where $\nu_E$ is the Liouville measure of the constant energy surface $\Sigma_E$ (Weyl law). \\

The expression at the right hand side of Eq.(\ref{promedio}) will be used for the theoretical
semiclassical analysis of $\langle N_B(E)\rangle$ but not as a recipe to implement a numerical
computation because it is difficult to properly sample
small velocities to generate the velocities distribution. However, we
are not concerned with the numerical evaluation of this distribution
but with the average of $N_B(E)$ which can be computed in different
ways. In Sec. \ref{bloch_basis} we describe a numerical method that allows to
compute $\langle N_B(E)\rangle$ accurately bypassing the computation of $P_\hbar(v,E)$. [See in
particular the last paragraph at the end of the aforementioned
section].\\

We now turn to obtain the leading order term of the semiclassical expansion of $\langle N_B(E)\rangle$. Asch and  Knauf \cite{Asch:1998p100} proved the asymptotic convergence of $\langle P_\hbar(v,E) \rangle$ towards the classical asymptotic velocities distribution $P_a(v,E)$, i.e., 
\begin{equation}
 \label{asch-knauf}
 P_\hbar(v,E) = P_a(v,E) \qquad \mbox{when}\quad
\hbar \rightarrow 0,
\end{equation}
that holds for test functions independent of $\hbar$, i.e. over intervals of order $\delta v \sim 1$, $\delta E \sim 1$, a coarse graining consistent with (\ref{promedio}). In the right-hand side of (\ref{asch-knauf}) $P_a(v,E)$ is the probability density of the asymptotic mean velocity
\begin{equation}
 \label{classical-asymptotic-velocity-dist}
 v_a=\lim_{t\rightarrow\infty} \frac{x(t)}{t}
\end{equation}
where $x(t)$ is the longitudinal position at time $t$ of a free particle in the waveguide whose initial condition was taken randomly with a uniform probability distribution on the surface $\Sigma_{E}$. When the classical dynamics is not purely ergodic, so there are some tori associated to ballistic trajectories in the stroboscopic Poincare section, the leading order term in the semiclassical limit of $N_B$ can be obtained just using equivalence (\ref{asch-knauf}) in the integrand of Eq.(\ref{promedio}). On the other hand, in case the dynamics is completely ergodic, $P_a(v,E)dv$ is a punctual measure at $v=0$ because $v_a=0$ almost surely. To obtain the leading semiclassical contribution in $\langle N_B(E)\rangle$, it is necessary to quantify more precisely how the quantum level velocity distribution
approaches the classical asymptotic velocity distribution as $\hbar \to 0$, that is, we need a refinement of (\ref{asch-knauf}).\\

When the classical dynamics is fully ergodic and the mixing rate is rapid enough, we expect that (almost) any initial ensemble of particles in the system will relax by diffusion, with a Gaussian density profile spreading according to $\langle x(t)^2 \rangle = D_E t $ for large times, with $D_E$ the diffusion coefficient for $\Sigma_{E}$. Hence, the mean velocity distribution at time $t$ is given by
\begin{equation}
 \label{P_t}
 P_t(v,E) = \frac{\nu_E}{\sqrt{2\pi}\bar{\sigma}_t} \exp{\left(-\frac{v^2}{2\bar{\sigma}_t^2}\right)}
\end{equation}
where $\bar{\sigma}_t^2 = D_E/t$. In the limit $t \rightarrow\infty$, $P_t(v,E)$ converges to a punctual measure at $v=0$. 
It has been shown \cite{Eckhardt:1995p110}, using Gutzwiller trace formula and also with RMT arguments, that in systems with hyperbolic classical dynamics the variance 
of a quantum operator in the semiclassical limit is equal to the variance of the associated classical observable at time $t=t_H/g$, with $t_H =h/\triangle E=\nu_E/h$ the Heisenberg time of the unit cell and $g$ a factor depending on the anti-unitary symmetries of the system. For the billiard we study in Sec. \ref{Sec.3} the factor is $g=2$ due to the transversal mirror reflection symmetry of the unit cell \cite{Berry} (see the discussion in Sec. \ref{Sec.5}). Then, we have a relation between the variance $\sigma_\hbar^2=\langle v_{n,\theta}^2\rangle$ of the quantum level velocity and the variance of the classical mean velocity $\bar{\sigma}_t$ 
\begin{equation}
 \label{eckhardt_equiv}
 \sigma_\hbar^2 = \frac{g D_E}{t_H} =\bar{\sigma}_{t_H/g}^2 \quad \mbox{for}\quad \hbar\rightarrow 0. 
\end{equation}
This relation between level velocity variance and the classical diffusion coefficient $D_E$ for the energy shell $\Sigma_E$
was studied in detail for
the kicked rotor \cite{arul} where the diffusion coefficient displays a nontrivial behavior as a parameter is varied. 
In a different context \cite{felipe} explores the relation between classical diffusion coefficient and quantum
evolution properties.\\

Inspired by Eq.(\ref{eckhardt_equiv}), Faure conjectured a higher order semiclassical equivalence between 
the averaged distribution of quantum velocities $\langle P_\hbar(v,E) \rangle$ and the classical mean velocity distribution at time $t=t_H/g$, i.e.,
\begin{equation}
 \label{faure_conjecture}
   P_\hbar(v,E) = P_{t=t_H/g}(v,E) \quad \mbox{when}\quad  \hbar \rightarrow 0 \, ,
\end{equation}
over widths of order $\delta E\sim 1$, $\,\delta v\sim \sqrt{h}$. This conjecture is based on result (\ref{eckhardt_equiv}) and the Quantum Ergodicity Theory \cite{Backer:1998p148}. 
It can also be seen as a consequence of the universality in the parametric variation of energy levels discovered by Simons and Altshuler \cite{Simons:1993p111}, as we discuss in Sec. \ref{Sec.5}.
Using (\ref{faure_conjecture}) in the integral of (\ref{promedio}) we have, for a purely diffusive waveguide, 
\begin{equation}
 \label{nb_k_diff}
 \langle N_B(E) \rangle= \frac{1}{\sqrt{h}}\sqrt{\frac{g\,\nu_E D_E}{2\pi L^2}} + o(\sqrt{1/h}).
\end{equation}
Equation (\ref{nb_k_diff}) was derived in \cite{Faure:2002p3} without the factor $g$
and checked numerically for the kicked Harper model. \\

Billiards are a particular class of systems with the simplifying property that their phase-space dynamics is the same on every energy shell except for a change of time scale. For two-dimensional billiards $\nu_E=2\pi A_c$ depends only on $A_c$, the area of the unit cell. Hence, the semiclassical limit $\hbar \to 0$ is equivalent to $k \to \infty$ due to the relation $\hbar^2k^2=2 E$. In $\Sigma_E$ the upper bound of the mean asymptotic speed is $\sqrt{2E}$. Considering the change of variables $v=\sqrt{2E}u$ we have (from Eqs. \ref{promedio} and \ref{asch-knauf}) $\langle N_B(E) \rangle = \frac{\sqrt{2E}}{h L} \int_0^1 u P_a(u) \,du $ with $P_a(u)$ the classical asymptotic velocity distribution in $\Sigma_{1/2}$ where the particles move with constant unit speed. 
Since $\sqrt{2E}=\hbar k$,
\begin{equation}
 \label{nb_quant}
\langle N_B(k) \rangle = \frac{k}{2\pi L} \int_0^1 v P_a(v) \,dv.
\end{equation}
Thus, for ballistic waveguides ($P_a(v)\neq \nu_E \delta(v)$)  the average number of propagating Bloch modes 
goes asymptotically for $k$ going to infinity as 
$\sim \mu_{bal} k/2\pi  + o(k)$
where $\mu_{bal}$
is the measure of ballistic trajectories in a stroboscopic cross section of the waveguide as defined in \cite{Faure:2002p3}. For diffusive billiard chains, the diffusion coefficient satisfies $D_E=|v|D_1$ with speed $|v|= \sqrt{2E}$ and $D_1$
the diffusion coefficient in the energy shell $\Sigma_{1/2}$ where particles move with unit speed. Then, using $|v|=\hbar k$ and $\nu_E=2\pi A_c$ in Eq.(\ref{nb_k_diff}) we have that the average number of propagating modes in a diffusive wave guide is
\begin{equation}
 \label{nb_kk_diff}
 \langle N_B(k) \rangle= \frac{\sqrt{g\,A_c D_1}}{\sqrt{2\pi} L} \sqrt{k} + o(\sqrt{k})
\end{equation}
In the following sections we will study how well this result applies to a version of the cosine waveguide, where many semi-classical results for quantum systems have been tested \cite{mmendoza, jmendez}.

\begin{figure}
 \centering
\includegraphics[width=0.95\columnwidth]{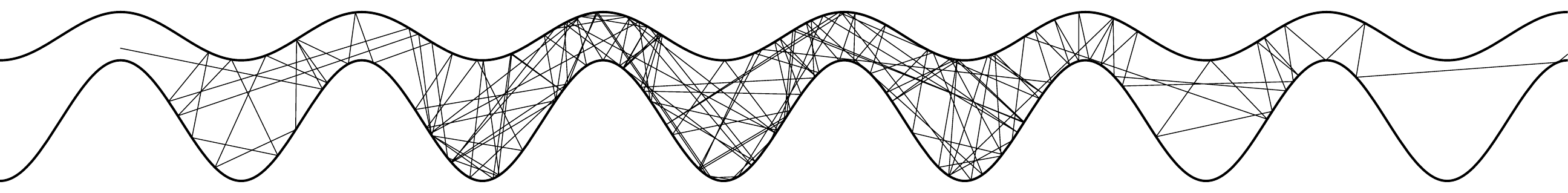}
 \caption{Portion of the infinite cosine waveguide with a trajectory's segment inside.}
 \label{fig-cosine-extended} 
\end{figure}

\section{Cosine billiard chain}
\label{Sec.3}
\subsection{Definition and classical dynamics}
\begin{figure}
 \centering
 \includegraphics[width=0.90\columnwidth]{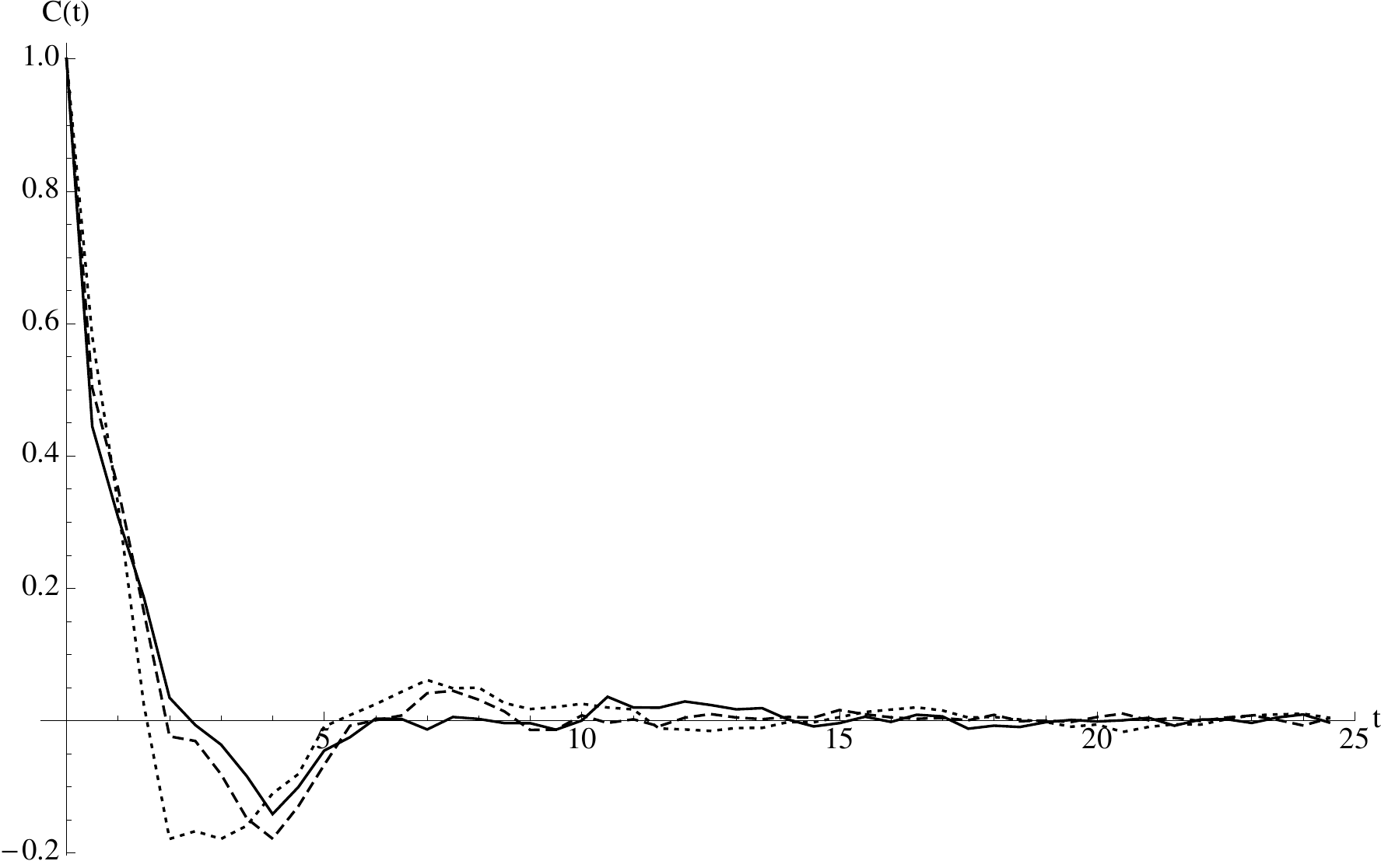}
 \caption{Velocity autocorrelation function $C(t)=\langle v_x(t)v_x(0)\rangle$ for an ensemble of 20000 particles in three billiards with $A_2=0.6,\,0.45,\,0.3$ (full, dashed and dotted lines respectively). The decay to noise level is quite fast in all cases, at around $t\sim15$ which is approximately 30 collisions.}
 \label{fig-vx-correlation} 
\end{figure}

\begin{figure}
 \centering
\includegraphics[width=0.95\columnwidth]{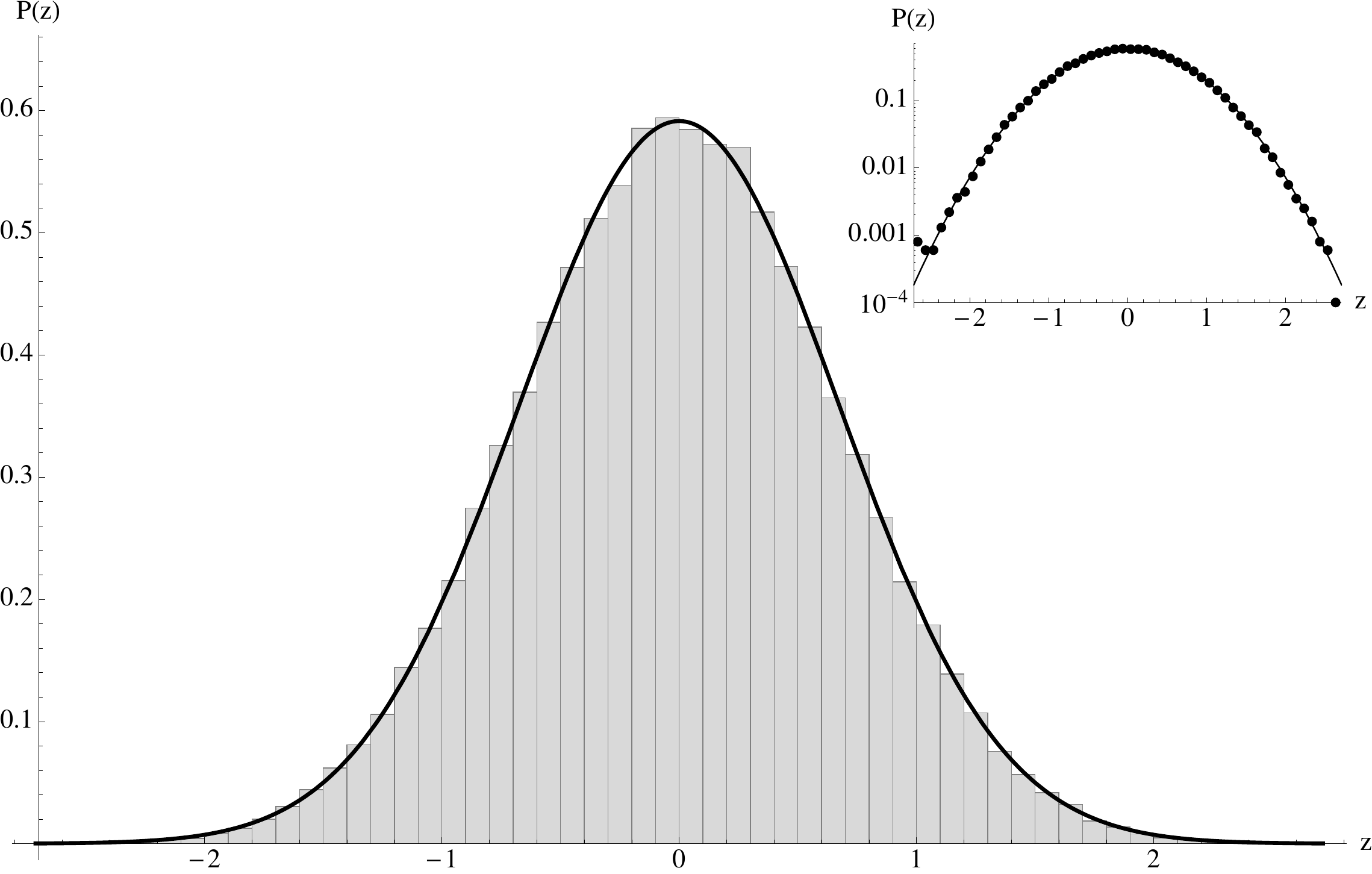}
 \caption{Histogram of the normalized displacement $\tilde{x}_t$ for a waveguide with $A_2=0.3$ at time $t=50000$ using an ensemble of $10^5$ initial conditions. The best Gaussian fit is plotted over it. The inset shows the same histogram (dots) in log-scale where convergence to a Gaussian (full line) is more evident to be achieved even deep in the tails.} 
 \label{fig-dist-z} 
\end{figure}

\begin{figure}
 \centering
\includegraphics[width=0.90\columnwidth]{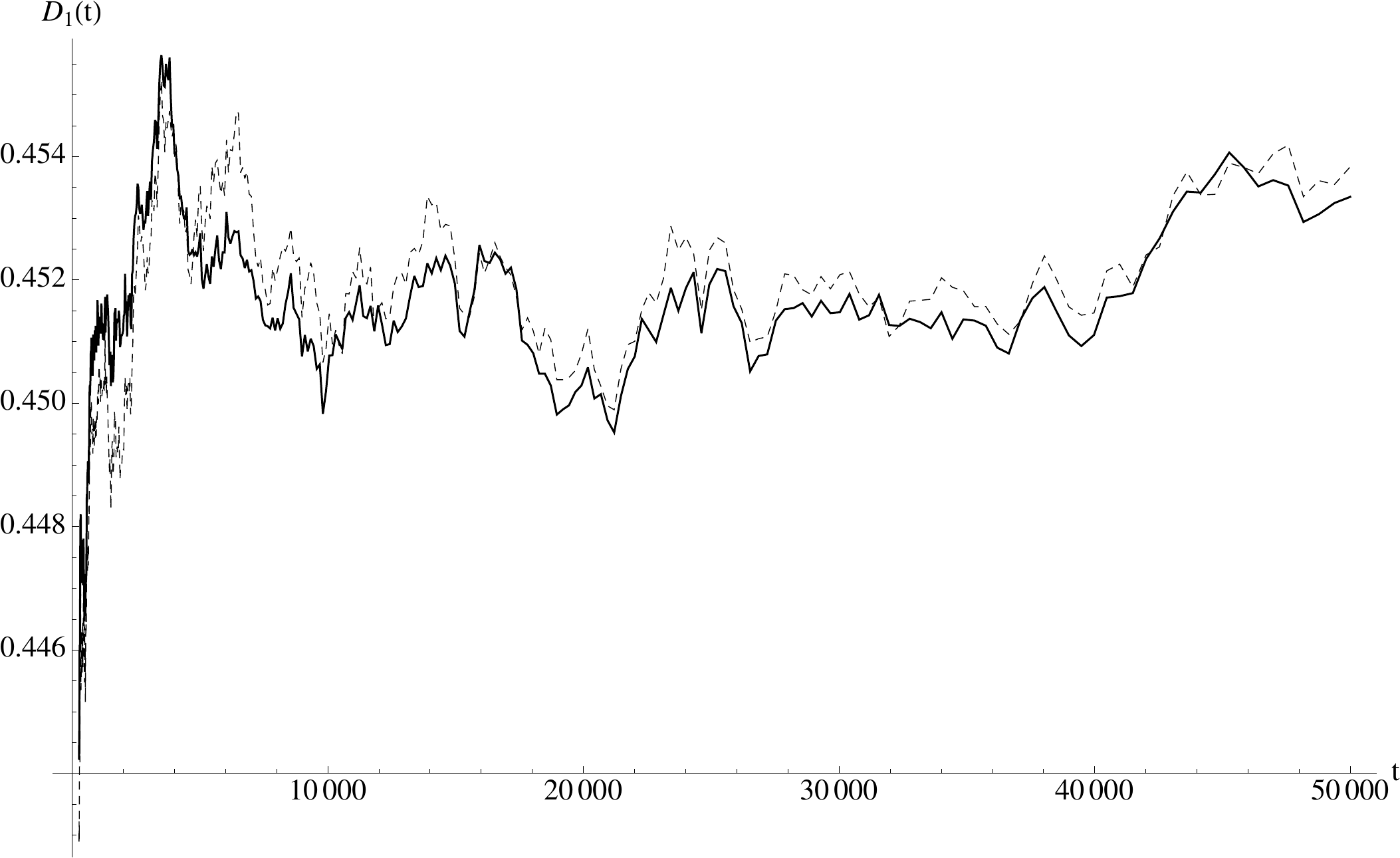}
 \caption{Finite time approximations of the diffusion coefficient $D_1(t) = \langle x_t^2 \rangle / t$ for the $A_2=0.3$ cosine waveguide calculated from the variance (second moment) of the sample's displacements (black line) and from the first moment using that $\frac{\pi}{2} \langle |\tilde{x}_t| \rangle^2 = \langle \tilde{x}_t^2 \rangle$, as expected for Gaussians, (dashed line) for an ensemble of $10^5$ initial conditions. To obtain the diffusion coefficient $D_1$ we average $D_1(t)$ for times after the transient relaxation regime. The diffusion coefficient obtained in this way from the first and second moments of $x_t$ are the same up to standard error.}
 \label{fig-moments-gauss} 
\end{figure}

We consider a periodic bidimensional waveguide with cosine-shaped hard walls  \cite{lunaacosta:1996}. The unit cell is defined for $-1\leq x\leq 1$ with the upper boundary given by $h_2(x)=1+ \frac{A_2}{2}(1+\cos{(\pi x)})$ and the lower boundary defined by $h_1(x)=\frac{1}{2}(1+\cos{(\pi x)})$. The waveguide is constructed as an infinite one-dimensional chain of these unit cells as shown in Fig.  \ref{fig-cosine-extended}. Such a billiard has finite horizon, i.e. it does not have unbounded collision-free trajectories. Note the aforementioned unit cell mirror symmetry in the transversal axis ($x=0$). We focus on a few values of $A_2$ in the range $0.3 \leq A_2 \leq 0.60,$ where we found that the dynamics is strongly chaotic. Very small tori are observed for $A_2=0.38$ and $0.45$ in the longitudinal Poincare section over $y=h_1(x)$. For the others values of $A_2$ considered the phase space looks fully ergodic to the naked eye. In all cases the connected ergodic component makes up at least about 95\% of phase-space volume and the most important property for our purposes,  exponential decay of the instantaneous velocity auto-correlation $\langle v_x(t)v_x(0)\rangle$, was observed for all values of $A_2$ explored (Fig.  \ref{fig-vx-correlation}). This implies that the system exhibits normal diffusion, i.e., that
\begin{equation}
 \label{rv}
 \tilde{x}= \lim_{t\rightarrow\infty} \frac{x(t)-x(0)}{\sqrt{t}} \equiv \lim_{t\rightarrow\infty} \tilde{x}_t
\end{equation}
is a stationary Gaussian random variable, with zero mean and variance $\langle \tilde{x}^2 \rangle = D_1$ the diffusion coefficient of the guide. In Fig.  \ref{fig-dist-z} we plot the distribution of the normalized displacement $\tilde{x}$ for $A_2=0.3$ showing remarkably good convergence to a normal distribution. This holds true for the other values of $A_2$ that we consider. Another way to support our conclusion that $\tilde{x}$ is normal-distributed and that 
there is convergence for at least the first two moments \cite{papulis} comes from the fact that
$ \frac{\pi}{2} \langle |\tilde{x}_t| \rangle^2 = \langle \tilde{x}_t^2 \rangle $
for large enough time as observed in Fig.  \ref{fig-moments-gauss}. 
Note that, for instance, an anomalous diffusive system with ergodic phase-space but infinite horizon trajectories could have convergence in distribution to a Gaussian for an appropriate normalized displacement but would fail to satisfy this relation between first and second moments for any finite time approximation of $\tilde{x}$ \cite{Armstead:2003p149}. 

\subsection{Bloch basis}\label{bloch_basis}
\begin{figure}
 \centering
 \includegraphics[width=0.90\columnwidth]{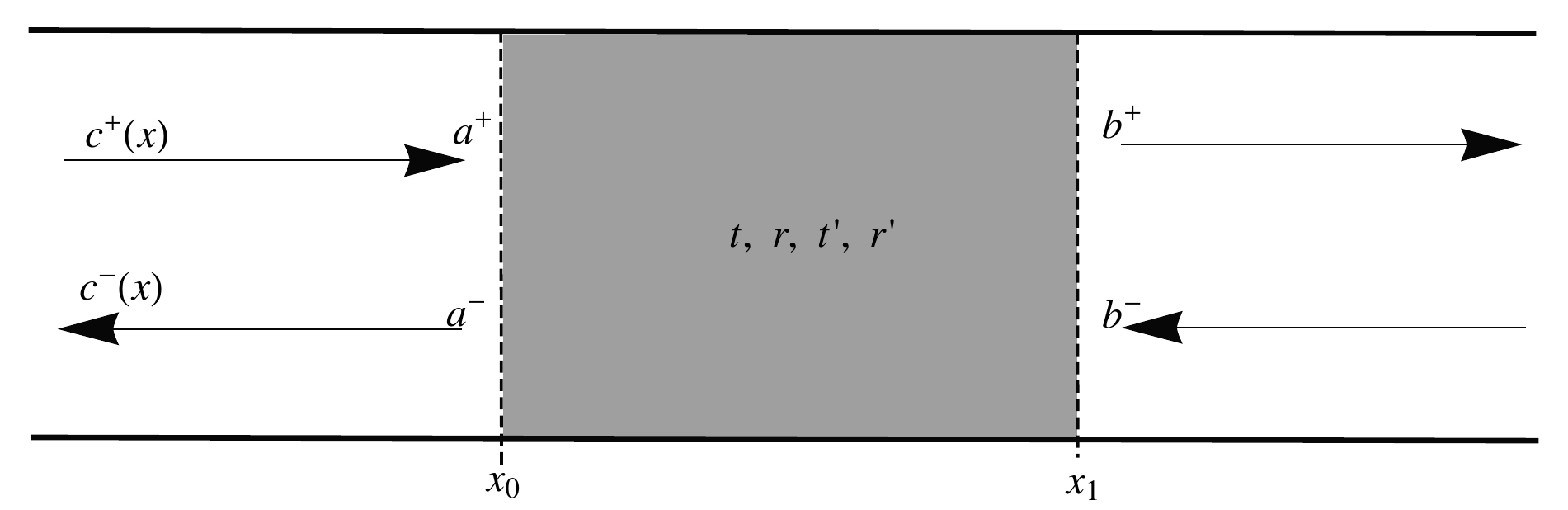}
 \caption{Illustration of a quasi-one-dimensional waveguide composed of one unit cell (gray) connected to two ideal leads going to infinity. The field (Eq. \ref{field}) at the unit cell's boundaries is given by $\left[ c_i^{+}(x_0) \;\; c_i^{-}(x_0)\right]=\left[ a_i^{+} \;\; a_i^{-}\right]$ and $\left[ c_i^{+}(x_1) \;\; c_i^{-}(x_1)\right]=\left[ b_i^{+} \;\; b_i^{-}\right]$. Solving the scattering problem in this geometry we can obtain the Bloch basis of the infinite periodic chain.}
 \label{fig-bloch-basis} 
\end{figure}

The propagating states of a periodic waveguide are given by the Bloch modes with unitary eigenvalues. The Bloch basis of a given waveguide can be defined by the solutions of a generalized eigenvalue problem as follows \cite{duclos:2009}.
Consider only one unit cell between two infinite plane leads. The wave-function can be written as
\begin{equation}\label{field}
\psi(x,y) = \sum_{i=1}^{\infty} \left( c_i^+(x) +  c_i^-(x)\right)\phi_i(x,y)
\end{equation}
where $i\in\mathbb{N}$ labels the local transverse modes $\phi_i(x,y)$ which satisfy the waveguide boundary conditions at each $x$. In the particular case we are considering (hard-wall billiard chain) $\phi_n(x,y)=\sqrt{\frac{2}{h(x)}}\sin\left(\frac{n\pi(y-h_1(x))}{h(x)}\right)$ with $h(x) = h_2(x) - h_1(x)$. Inserting Eq. (\ref{field}) in Eq. (\ref{helmholtz}), the wave equation becomes a system of ordinary differential equations for the $c^{\pm}_i(x)$ which after suitable transformation can efficiently determine transmission and reflection matrices \cite{Pagneux:2006,Pagneux:2009}. 
Let $a_i^+$ ($a_i^-$) be the right (left) going wave on the left lead and $b_i^+$ ($b_i^-$) the right (left) going wave on the right lead (identified with $c^{\pm}_i(x)$ evaluated at the left $x_l$ and right $x_r$ boundaries of the cell respectively) as illustrated in Fig.  \ref{fig-bloch-basis}. If $t$ and $r$ ($t'$ and $r'$) are the left-to-right (right-to-left) transmission and reflection matrices 
then outgoing waves are linked to ingoing waves by
\begin{equation}
 \label{eq:2}
 \begin{array}{ll}
   t a^+ + r' b^- & = b^+\\
   r a^+ + t' b^- & = a^-
 \end{array}
\end{equation}
We can recast this (infinite) system of equations as 
\begin{equation}
M_1 \left[
\begin{array}{c}
 a^+ \\ a^- 
\end{array}\right]
=M_2 \left[
\begin{array}{c}
 b^+ \\ b^-
\end{array}\right]
\end{equation}
where
\begin{equation}
 \label{gen_eigenvalue_prob}
   M_1 =\left(\begin{array}{cc}
   t & 0\\
   -r & I\\
 \end{array}\right)
\qquad
M_2 = \left(\begin{array}{cc}
   I & -r'\\
   0 & t'\\
 \end{array}\right)
\end{equation}
Now, we impose Bloch condition, namely that the wavefunction is equal on both lead boundaries up to a complex factor $\lambda$,
\begin{equation}
 \label{bloch_def}
\left[
\begin{array}{c}
 a^+ \\ a^- 
\end{array}\right]
= \lambda \left[
\begin{array}{c}
 b^+ \\ b^-
\end{array}\right]
\end{equation}
where $\lambda=\exp(i\theta)=\exp(iqL)$. Then, the Bloch basis is defined as the set of modes that satisfy (\ref{bloch_def}), i.e. they are the solutions $\mathbf{v}_n$ of the unit cell's generalized transfer matrix eigenvalue problem 
\begin{equation}
 \label{M_egvp}
 M_1 \mathbf{v}_n = \lambda_n M_2 \mathbf{v}_n
\end{equation}
The propagating Bloch modes are such that $|\lambda_n| = 1$; the other modes are evanescent and decay exponentially in the direction of propagation. In practice, this infinite system of equations must be truncated to a dimension at least equal to the number of propagating Fourier modes in the leads but in most cases it is necessary to include some evanescent Fourier modes to describe the field adequately. Note that the transmission and reflection matrices used in (\ref{gen_eigenvalue_prob}) are evaluated at the (right or left) entry boundary of the unit cell and carry information from these decaying modes in the near field, which in general can couple between unit cells in the chain. Hence, these matrices do not satisfy the usual (energy) flux conservation relations coming from the unitarity of the scattering matrix in the far field \cite{Pagneux:2004}. Moreover, including evanescent modes brings about infinitely fast decaying states (i.e. null eigenvalues $\lambda_n$) \cite{Wu:1994p33} making the matrices (\ref{gen_eigenvalue_prob}) non-invertible. In this case the transfer matrix $M=M_2^{-1}M_1$ cannot be defined. Only when the geometry of the unit cell is such that the coupling between evanescent modes is negligible it is possible to truncate $t$, $r$, $t'$ and $r'$ to the far field non-decaying modes and define the transfer matrix.

Note that our method computes $\theta$ as a function of $k$, i.e., we fix the energy $E$ (or the wavenumber $k$) and then
we obtain the set of Bloch quasi-momentum $\theta_n$ for which a propagative solution exists. With this method we can 
accurately compute the energy average at the left hand side of Eq.(\ref{promedio}).


\section{Numerical Results}\label{section-numerics}

We now study numerically the semiclassical behavior of $\langle N_B(k)\rangle$ for the cosine billiard chain. In order to compute the transmission and reflection matrices we employ the admittance multi-modal method described in \cite{Pagneux:2009}, where, for each value of $k$, the field is described using the local transverse modes (Fourier basis) of the waveguide truncated to a finite number of channels, as described in Sec. \ref{bloch_basis}. The number of propagating modes $N_B(k)$ is obtained by counting the solutions of Eq.(\ref{M_egvp}) with $|\lambda_n| = 1$ and positive group velocity. As a function of $k$, the above number fluctuates at scales $\Delta k = 2\pi/(kA_c)$ \footnote{$\Delta k$ follows from the Weyl law $N=k^2 A_c/(4\pi)$ }. We consider an average of the form 
\begin{equation}
 \label{nb_av}
 \langle N_B(k) \rangle_r = \int N_B(k') f_r(k-k')\, dk'
\end{equation}
where $f_r(k)$ is a positive function of unit norm in $dk$ with compact support in an interval of length $r \Delta k$ with $r\gg 1$. The dependence of $\langle N_B(k) \rangle_r$ in the width of $f_r(k)$ is briefly discussed at the end of this section but for what follows we drop explicit reference to $r$. \\ 

\begin{figure}
 \centering
\includegraphics[width=0.95\columnwidth]{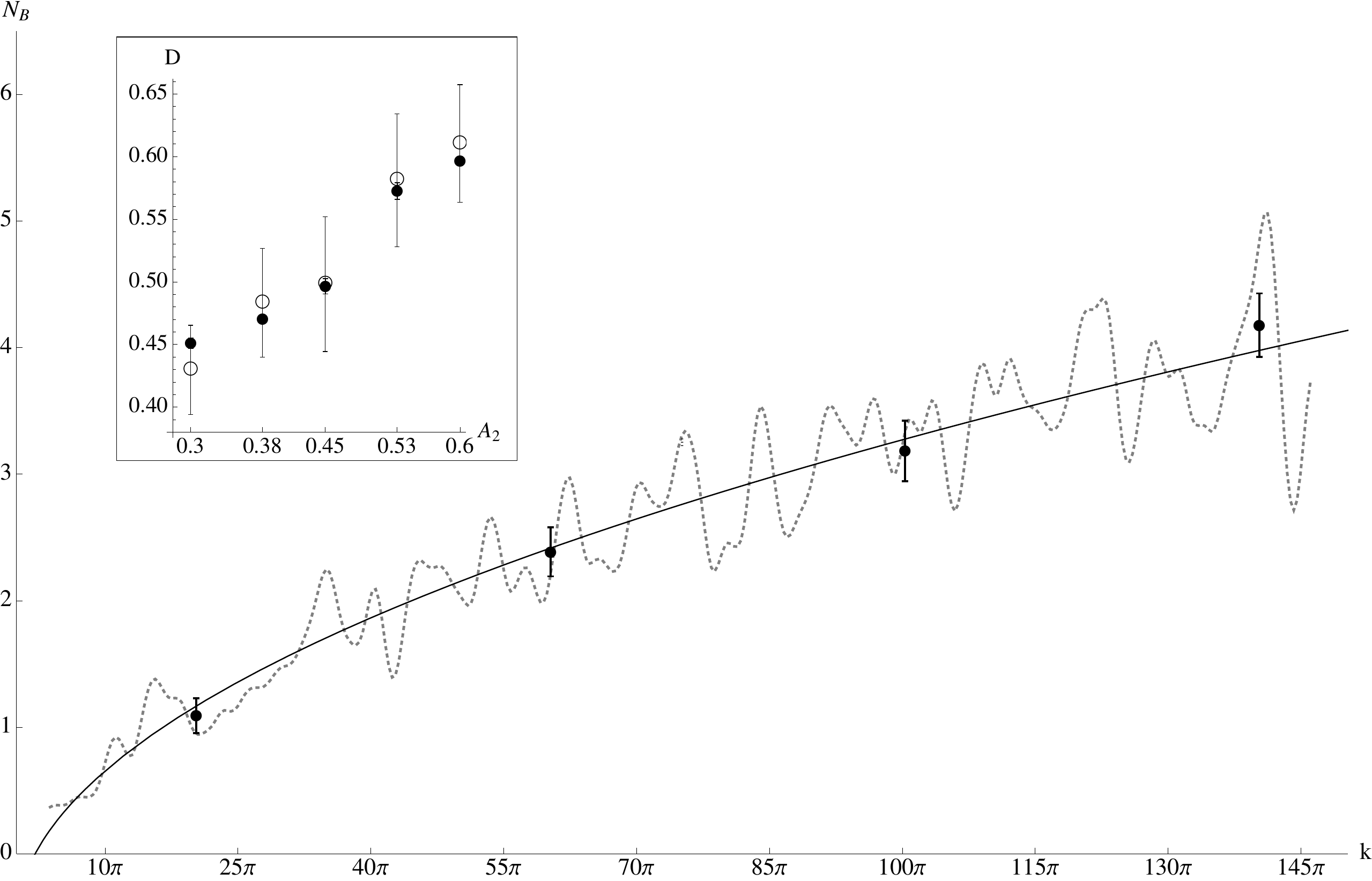}
 \caption{The dashed line is the gaussian smoothed moving average (with standard deviation $\sim \pi$) of the number of propagating Bloch modes $N_B(k)$ in the cosine billiard chain with $A_2=0.3$ for a uniform $k$ sampling with increments $\delta k =0.37147\pi$. The expected semiclassical result for $\langle N_B(k)\rangle$, plotted in full line, shows good agreement with the moving average. The four black points (with their associated standard error) were computed for the same geometry using a uniform sampling of 153 wavenumbers in an interval $300$ mean level spacings wide, which constitutes a better statistics than the moving average (a sampling at least 15 times more dense). In the inset, the diffusion coefficient calculated classically (black dots) and the one obtained from the best fit following  Eq. (\ref{nb_kk_diff}) for the quantum computation of $N_B(k)$ (circles) for five values of $A_2$ is shown. In all cases, a good agreement with the expected semiclassical behaviour Eq. (\ref{nb_kk_diff}) is observed.}
\label{fig-nb_k_0.30} 
\end{figure}

In Fig. \ref{fig-nb_k_0.30} we show $\langle N_B(k)\rangle$ for $k\in [\pi, 150\pi]$, for $A_2=0.3$. For this $A_2$ value, the phase-space does not show noticeable tori. In the $k$ interval explored the number of open transverse modes in the Fourier basis ranges from 1 up to 150 (100 evanescent modes were used for all wavenumbers). The dashed line shows a gaussian smoothed moving average with variance $\pi^2$ of $N_B(k)$ for a uniform sampling of wavenumbers with spacing $\delta k \sim 0.37\pi$, which is bigger than the mean level spacing in the whole $k$ range. The full line is the expected semiclassical result (\ref{nb_kk_diff}) adjusted with an additive constant. This constant appears because in the system under study $\langle N_B(k^*) \rangle= 0$ for some $k^*>0$ 
and we are not sufficiently deep in the semiclassical regime for its relative value to be negligible. $N_B(k)$ fluctuates very rapidly with $k$ as expected but the agreement between the moving average and the semiclassically predicted curve is quite good in the whole wavenumber interval explored.  We observe similar good results for other configurations of the unit cell with $0.6\geq A_2\geq 0.3$ even when there are some small tori in the transverse Poincare sections. This is not surprising because the classical dynamics in those systems is also strongly mixing and for waves to resolve such small structures wavenumbers of order $k\gtrsim 2000 \pi$ would be needed, a fairly large value compared to the range considered. In the inset of Fig.  \ref{fig-nb_k_0.30} we show the diffusion coefficient calculated classically (from Montecarlo simulations) and also the one obtained from the best fit following (\ref{nb_kk_diff}) for the quantum computation of $N_B(k)$ for five values of $A_2$. \\

\begin{figure}
 \centering 
 \includegraphics[width=0.95\columnwidth]{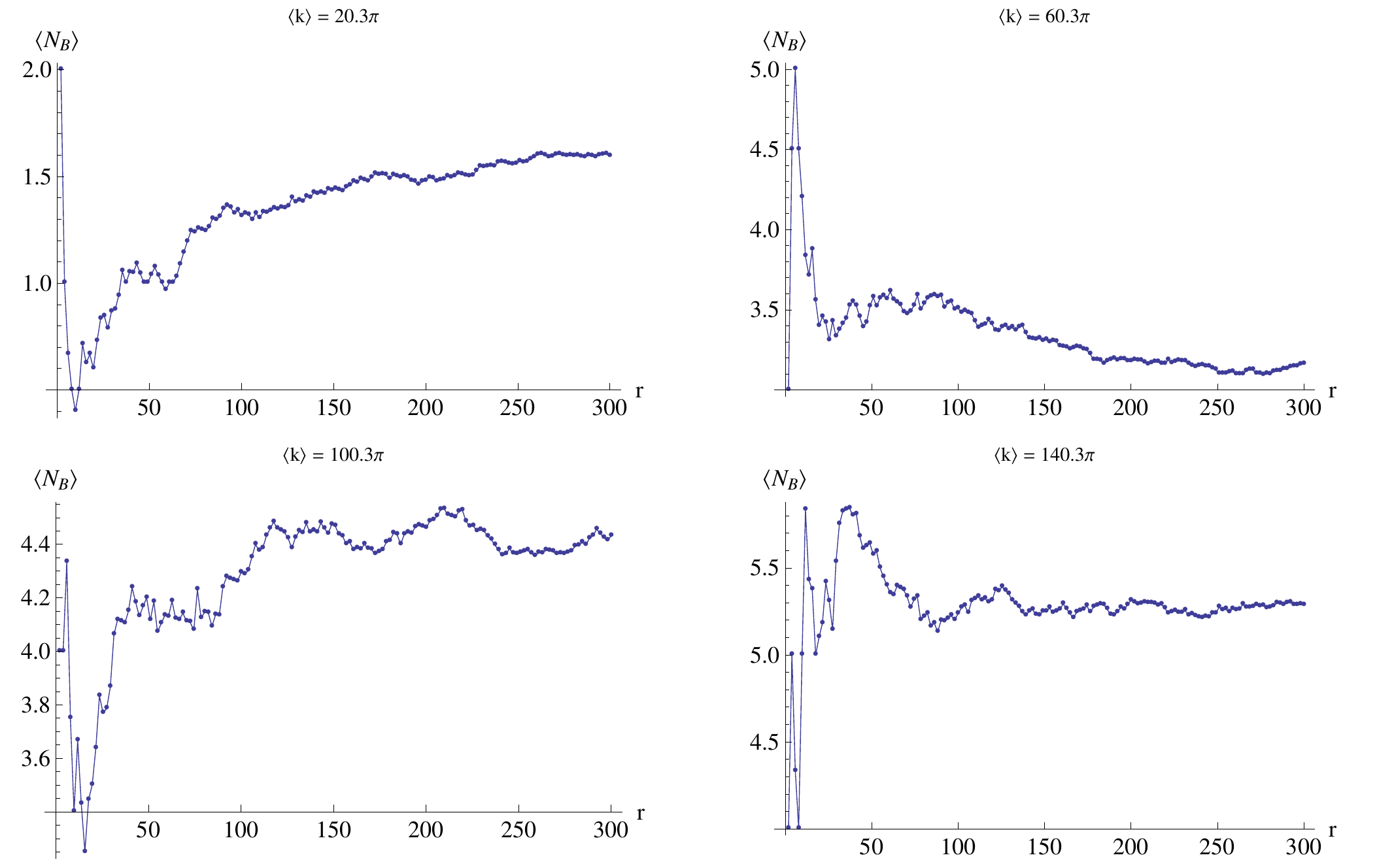}
 \caption{$\langle N_B(k)\rangle_r$ as a function of $r$ for four energies in the $A_2=0.3$ cosine billiard. A uniform sampling of 153 wavenumbers distributed in a 300 mean-level-spacing wide interval was used. In all cases after $r \sim 200 $ the average fluctuations are notably reduced and the result for $\langle N_B(k)\rangle$ is in agreement with the expected semiclassical result (see Fig.  \ref{fig-nb_k_0.30}) indicating that this is a good smoothing window to eliminate the spectrum fluctuations.}
 \label{fig-bb_fat_points_convergence} 
\end{figure}

As mentioned in the definition Eq.(\ref{nb_av}) of the mean number of propagating Bloch modes, the width of interval $r\Delta k$ over which we take the average is arbitrary and it is only assumed to be much larger that the mean level spacing $\Delta k$ and classically small. In order to find out how relevant is this choice we calculate $\langle N_B(k)\rangle_r$ for four different values of $k$ as a function of $r\in [2, 300]$ using a uniform $k$ sampling of 153 points around the four wavenumbers. In Fig.  \ref{fig-bb_fat_points_convergence} we see that, after a transient phase in $r\sim 200$, the value of the average reaches an approximately stable region in the four cases.  The four black dots in Fig.  \ref{fig-nb_k_0.30} show this result for $r=300$, also agreeing nicely with the expected semiclassical curve.

We end this section remarking how close to the semiclassical regime the presented results  are. In our calculations $\lambda/W$ (the wavelength in units of the waveguide width) is as small as $0.012$.  We note that the calculations in \cite{Dittrich:1997p106}, which are also compared with semiclassical results for periodic billiards, were computed with the parameter $\lambda/W \sim 0.12$.

\section{Discussion and Conclusions}
\label{Sec.5}

In section \ref{section-bloch_modes} we closely followed Faure derivation of Eq.(\ref{nb_k_diff}). Here we provide an alternative argument allowing to understand better the connection with \cite{Simons:1993p111,Dittrich:1998p118}. Proceeding as in Sec. \ref{section-bloch_modes}, we define the probability distribution of level velocity in natural units $w$
\begin{equation}
p(w,E)=\frac{1}{2\pi}\sum_n \int_{-\pi}^{\pi} d\theta\, \delta(E-E_{n,\theta}) \delta(w-\frac{dE_{n,\theta}}{d\theta}) 
\end{equation}
so that 
\begin{equation}
     N_B(E) = 2\pi \int_0^\infty w p(w,E)\,dw.
     \label{26}
\end{equation}
Then, as was noted in \cite{zirnbauer}, the distribution of level velocities satisfies $p(w,E)=\lim_{\phi \to 0}\phi K(w\phi,\phi)$ with $K(\Omega,\phi)$
the autocorrelation function
\begin{equation}
K(\Omega,\phi)=\langle \rho(E+\Omega,\theta+\phi)\rho(E,\theta)\rangle_{E,\theta}-\bar{\rho}_E^2
\end{equation}
of the energy density $\rho(E,\theta)=\sum_n \delta(E-E_{n,\theta}).$ The average is over $\theta$ and over a classically small energy interval around $E$ and $\bar{\rho}_E$ denotes this average for $\rho(E,\theta)$. According to Simons and Altshuler $k(\omega,\chi)=K(\Omega,\phi)/\bar{\rho}_E^2$ is a universal function where $\omega=\bar{\rho}_E\Omega $ and $\chi=\sqrt{C(0)}\phi$ with $C(0)=\bar{\rho}_E^2\langle \left(\frac{dE_{n,\theta}}{d\theta}\right)^2\rangle.$ They showed this result for disordered systems and verified numerically its validity for chaotic systems in the semiclassical limit. It follows from their expression for $k(\omega,\chi)$ that the probability distribution of level velocity $p(w,E)$ is a Gaussian with variance $C(0)/\bar{\rho}_E^2$. Thus, from Eq.(\ref{26}), we obtain  $N_B(E)=\sqrt{2\pi C(0)}.$ The constant $C(0)$ is system dependent and must be computed, for example, semiclassically. In our case it is simply related to the average of the square of the quantum velocity $C(0)=\frac{\bar{\rho}_E^2\hbar^2}{L^2}\langle v_{n,\theta}^2\rangle$ that is given in Eq.(\ref{eckhardt_equiv}), leading to $N_B(E)=\sqrt{\frac{1}{h}\frac{g \nu_E D_E}{2\pi L^2}}$. Thus, Faure's result Eq.(\ref{nb_k_diff}) can be seen as a consequence of universality in the parametric level correlations. The validity of the universal correlation function for periodic systems was studied in \cite{Dittrich:1998p118} by Dittrich et al. in the semiclassical limit. They computed $K(\Omega,\phi)$ by Fourier transforming their semiclassical expression for the form factor identifying a $C(0)$ that agrees with the previous result.\\

The universal correlation function $k(\omega,\chi)$ is different for GOE and GUE. For periodic systems, when $\chi$ plays the role of the Bloch parameter, time reversal symmetry $T$ is broken for all $\chi$ besides three exceptional points \cite{Dittrich:1997p106}. Thus, when no further anti-unitary symmetries exist, $k(\omega,\chi)$ moves between GOE and GUE at different parts of the band. In our case, the unit cell is invariant under the operator $S_x$ that transforms $x \to -x$ (mirror reflection along the transversal direction). So, even though the unit-cell hamiltonian does not commute with $S_x$ nor with $T$, it does commute with the operator $T S_x$ which is anti-unitary. Then, as $\chi$ is varied, the system always belongs to GOE \cite{Berry}.\\

As a final comment, we note that Eq.(\ref{nb_quant}) applies also to classical waves with dispersion relation $\omega=c k$ propagating in a wave-guide with speed $c=1$. We can restore $c$ by noticing that the distribution of group velocities $v_{n,\theta}=L\frac{d\omega_{n,\theta}}{d\theta}$
is related to our previous distribution by $p_k(v)=\frac{1}{c}P_k(\frac{v}{c})$. With this new $p_k(v)$ the formula becomes
\begin{equation}
 N_B(k)  =\frac{k}{2\pi c L} \int_0^\infty v p_k(v) \,dv
\end{equation}
with the normalization $\int dv p_k(v)=\int \frac{dv}{c} P_k(\frac{v}{c})=2\pi A_c$ for the group velocity distribution.
The result for the number of propagating modes also applies for these classical waves.\\

To conclude, we have studied the agreement between numerical computations and a semiclassical approximation of the number of propagating modes in a diffusive periodic waveguide. We also showed the relation between $\langle N_B(E)\rangle$ and previously studied spectral properties of periodic systems.  The semiclassical results used in our analysis is previous to the discovery of how to include non-diagonal contributions to correlation functions. Nevertheless it was noted \cite{Eckhardt:1995p110,arul} that their role in the computation of the fluctuations in velocity should be very limited. Indeed, besides the average justified by the rapid phases variation, the observable amplitude (the velocity in this case) fluctuates around its null average, further justifying the diagonal approximation.

\begin{acknowledgments}
The authors acknowledge support from the Scientific Computing Advanced Training (SCAT) project through EuropeAid contract II-0537-FC-FA (http://www.scat-alfa.eu) and from ECOS grant No. C09E07. They also thank F. Lund for careful reading of the manuscript.  
FB acknowledges financial support from the Vicerrector\'ia de Investigaci\'on y desarrollo de la Universidad de Chile (ENL 09/04).
JZ thanks a Conicyt Ph.D. fellowship. 
\end{acknowledgments}



\end{document}